\documentclass[aip,cha,reprint,numerical]{revtex4-1}
\usepackage {CJK}
\usepackage{amsmath}
\usepackage{amssymb}
\usepackage{graphicx}
\usepackage{color}

\begin{document}
\begin {CJK*} {GB} { } 

\title{Collective mode reductions for populations of coupled noisy oscillators}

\author{Denis S.\ Goldobin}
\affiliation{Institute of Continuous Media Mechanics, UB RAS, Academician Korolev Street 1,
 614013 Perm, Russia}
\affiliation{Department of Theoretical Physics, Perm State University, Bukirev Street 15,
 614990 Perm, Russia}
\author{Irina V.\ Tyulkina}
\affiliation{Department of Theoretical Physics, Perm State University, Bukirev Street 15,
 614990 Perm, Russia}
\author{Lyudmila S.\ Klimenko}
\affiliation{Institute of Continuous Media Mechanics, UB RAS, Academician Korolev Street 1,
 614013 Perm, Russia}
\affiliation{Department of Theoretical Physics, Perm State University, Bukirev Street 15,
 614990 Perm, Russia}
\author{Arkady Pikovsky}
\affiliation{Institute for Physics and Astronomy,
University of Potsdam, Karl-Liebknecht-Strasse 24/25, 14476 Potsdam-Golm, Germany}
\affiliation{Department of Control Theory, Nizhny Novgorod State University,
Gagarin Avenue 23, 606950 Nizhny Novgorod, Russia}
\date{\today}

\begin{abstract}
We analyze accuracy of different low-dimensional reductions
of the collective dynamics in large
populations of coupled phase oscillators with intrinsic noise.
Three approximations are considered: (i)~the Ott-Antonsen ansatz,
(ii)~the Gaussian ansatz,
and (iii)~a two-cumulant truncation of the circular cumulant
representation of the original system's dynamics. For the
latter we suggest a closure, which makes the truncation, for small
noise, a rigorous first-order correction to the Ott-Antonsen ansatz, and
simultaneously is a generalization of the Gaussian ansatz.
The Kuramoto model with intrinsic noise, and the population
of identical noisy active rotators in excitable states with
the Kuramoto-type coupling, are considered as examples to test
validity of these approximations.
For all considered cases, the Gaussian ansatz is found
to be more accurate than the Ott-Antonsen one for
high-synchrony states only. The two-cumulant approximation
is always superior to both other approximations.
\end{abstract}

\maketitle

\end{CJK*}

\begin{quotation}
Synchrony of large ensembles of coupled elements can be characterised
by the order parameters -- the mean fields. Quite
often the evolution of these collective variables is surprisingly
simple, what makes a description with only a few order parameters
feasible.
Thus, one tries to construct accurate closed low-dimensional
mathematical models for the dynamics of the first few order parameters.
These models represent useful tools for gaining insight
into the underlaying mechanisms of some more
sophisticated collective phenomena: for example, one describes
coupled populations by virtue of coupled equations for the relevant order
parameters.
A regular approach to the construction of closed
low-dimensional systems is also beneficial for
dealing with phenomena, which are beyond the applicability scope
of these models; for instance, with such an approach, one can
determine constrains on clustering in populations.
There are two prominent types of situations, where the low-dimensional
models can be constructed: (i)~for a certain class of ideal
paradigmatic systems of coupled phase oscillators,
the Ott-Antonsen ansatz yields an
exact equation for the main order parameter;
(ii)~the Gaussian approximation for the probability density of the phases,
also yielding a low-dimensional closure,
is frequently quite
accurate. In this paper, we compare applications of these two model
reductions for situations, where neither of them is perfectly
accurate. Furthermore, we construct a new reduction approach which practically
works as a first-order
correction to the best of the two basic approximations.
\end{quotation}

\section{Introduction}
Models of globally coupled oscillators are relevant for many
applications in physics, engineering, living and social
systems~\cite{Pikovsky-Rosenblum-Kurths-2001-2003,Strogatz-2003,%
Filatrella-Nielsen-Pedersen-2008,Winfree-2001,Acebron-etal-2005,Pikovsky-Rosenblum-15}.
The main effect here is synchronization, i.e.\ appearance of a nontrivial
mean field due to interactions. This effect can be viewed as
a nonequilibrium phase transtion, where the
appearing ordered synchronized state
is described by a set of order parameters. The famous
Kuramoto model of coupled phase oscillators is a paradigmatic
example for the synchronization transition, it is completely solvable
in the thermodynamic limit of an infinite population. The properties
of the transition are also quite well understood if additionally
to the coupling, the oscillators are subject to independent noise terms.

A description of globally coupled noisy oscillators can be reduced,
in the thermodynamic limit of large ensemble, to a
nonlinear Fokker-Planck equation
(or to a Liouville equation in the noiseless case), which is a
system with an infinite number of degrees of freedom. If one wants
not simply find the stationary solutions, but to follow the evolution of
the distributions, the problem of the reduction of the
infinite-dimensional system to several essential
degrees of freedom arises. This closure problem is in the focus of this paper.
We will discuss and compare three variants of the reduction to a
few global
modes: (i) the Ott-Antonsen ansatz~\cite{Ott-Antonsen-2008},
(ii) the Gaussian ansatz, recently
considered by Hannay et al.~\cite{Hannay-Forger-Booth-2018} on the basis
of previous
works~\cite{Zaks-etal-2003,Sonnenschein-Schimansky-Geier-2013,Sonnenschein-etal-2015},
and (iii) the circular cumulant approach
suggested in Ref.~\cite{Tyulkina-etal-2018}. Neither of these approaches is
exact for a population of coupled noisy oscillators,
but they provide quite good approximations of the observed regimes.
We will compare their accuracy
for different ranges of parameters.

\section{Basic Models}

Our basic model is a population of phase oscillators $\varphi_k(t)$
with intrinsic noise:
\begin{equation}
\dot\varphi_k=\omega_k+\mathrm{Im}(2h(t)e^{-i\varphi_k})+\sqrt{D}\eta_k(t)\,.
\label{eq:phase}
\end{equation}
Here natural frequencies $\omega_k$ have a Lorentzian (Cauchy)
distribution $g(\omega)=\gamma/[\pi((\omega-\omega_0)^2+\gamma^2)]$,
and $\gamma$ is the distribution half-width. Parameter
 $D$ is the noise intensity, terms
$\eta_k$ are independent normalized white Gaussian random forces:
$\langle\eta_k(t)\eta_m(t')\rangle=2\delta_{km}\delta(t-t')$,
$\langle\eta_k\rangle=0$. The coupling is determined by the
complex force
$h$, common for all oscillators.
For the Kuramoto setup, this force is proportional to the mean field
which is just the Kuramoto order parameter of the population
\[
h=\frac{K}{2} Z_1,\qquad Z_1=\langle e^{i\varphi}\rangle=\frac{1}{N}\sum_{j=1}^N
e^{i\varphi_j}\;.
\]
The Kuramoto model for noisy oscillators thus
reads
\begin{equation}
\dot\varphi_k=\omega_k+\frac{K}{N}\sum_{j=1}^{N}\sin(\varphi_j-\varphi_k)+\sqrt{D}\eta_k(t)\,.
\label{eq:KM}
\end{equation}
With a slight modification of the common force $h$, namely
\[
h=\frac{a}{2}+\frac{K}{2} Z_1\,,
\]
one obtains the equations for a population of  noisy active rotators with the
Kuramoto-type coupling, treated in Ref.~\cite{Zaks-etal-2003}:
\begin{equation}
\dot\varphi_k=\omega_k-a\sin\varphi_k+\frac{K}{N}\sum_{j=1}^{N}
\sin(\varphi_j-\varphi_k)+\sqrt{D}\eta_k(t)\,.
\label{eq:ARM}
\end{equation}
Models \eqref{eq:KM} and \eqref{eq:ARM} are the basic systems we consider below.

\section{Mode equations}
In the thermodynamic limit $N\to\infty$,
starting from the Langevin equations
\eqref{eq:phase}, one can write for the distribution
density of the subpopulation of the
oscillators with natural frequency $\omega$ the Fokker-Planck equation
\begin{equation}
\frac{\partial w(\varphi,t|\omega)}{\partial t}=
-\frac{\partial}{\partial\varphi}\left[\mathrm{Im}(2h(t)e^{-i\varphi}) w\right]
+D\frac{\partial^2}{\partial \varphi^2}w\,.
\label{eq:fpf}
\end{equation}
This equation can be rewritten as an infinite system for the complex amplitudes
of the Fourier modes
\[
z_m=\int_{-\pi}^\pi \mathrm{d}\varphi\, w(\varphi,t|\omega) e^{i m \varphi}
\]
of the density
$w(\varphi,t|\omega)=(2\pi)^{-1}\sum_m z_m(t,\omega)e^{-i m\varphi}$:
\begin{equation}
\dot{z}_n=n i\omega z_n+n h z_{n-1}-n h^\ast z_{n+1}-n^2Dz_n\,.
\label{eq:modes}
\end{equation}
The quantities $z_n$ are the local order parameters at a given frequency,
the global Kuramoto-Daido order parameters are obtained by
the additional averaging over the distribution of the natural frequencies:
\begin{equation}
Z_n=\int\mathrm{d}\omega\, g(\omega) z_n\,.
\label{eq:int}
\end{equation}
The main order parameter $Z_1$ is employed in the
definition of the forces $h$ in the two models we study in this paper.
Below we consider only the Lorentzian distribution $g$ and
adopt the assumption by Ott and Antonsen~\cite{Ott-Antonsen-2008}
on the analyticity of $z_n(t,\omega)$ as a function of
complex $\omega$ in the upper half-plane. This allows for
calculating the global Kuramoto-Daido order parameters via residues as
\[
Z_n=z_n(\omega_0+i\gamma)\,.
\]
In this way, one obtains an infinite
system of equations for $Z_n$ (which are in fact
moments $\left\langle \left(e^{i\varphi}\right)^n\right\rangle$ of the complex
observable $e^{i\varphi}$) with $Z_0\equiv 1$:
\begin{equation}
\dot{Z}_n=n(i\omega_0-\gamma)Z_n+nhZ_{n-1}-nh^\ast Z_{n+1}-n^2DZ_n\,.
\label{eq:daido}
\end{equation}

\section{Finite-dimensional reductions}
As we discussed above, it is desirable to reduce, at least
approximately, the infinite system~\eqref{eq:daido} to a
finite-dimensional one, and, in what follows, we discuss three ways
to accomplish this.

\subsection{Ott-Antonsen reduction}
Here one assumes, following Ref.~\cite{Ott-Antonsen-2008},
 that all the higher order parameters can be expressed via the first one
 according to
\begin{equation}
Z_n=(Z_1)^n.
\label{eq:OAsc}
\end{equation}
This reduces the system \eqref{eq:daido} to just one equation
\begin{equation}
\dot{Z}_1=(i\omega_0-\gamma)Z_1+h-h^\ast Z_1^2-DZ_1\,.
\label{eq:Z1OA}
\end{equation}
The Ott-Antonsen (OA) reduction works exactly for $D=0$, where it
defines the so-called OA invariant manifold. This
manifold corresponds to the probability density being the
wrapped Cauchy distribution of the phases.

\subsection{Gaussian reduction}
Recently, on the basis of
the analysis of some experimental data, another
representation of the higher order parameters through
the first one was suggested~\cite{Hannay-Forger-Booth-2018}:
\begin{equation}
Z_m=|Z_1|^{m^2-m}Z_1^m\,.
\label{eq:gausans}
\end{equation}
Equivalently, if we introduce the amplitude and
the argument of the Kuramoto order
parameter $Z_1=R_1e^{i\psi}$, with $R_1=\exp[-s^2/2]$,
we can rewrite \eqref{eq:gausans} as
\begin{equation}
Z_m= R_1^{m^2}e^{im\psi}=e^{-\frac{1}{2}m^2 s^2}e^{im\psi}.
\label{eq:gauss}
\end{equation}
This relation means that the corresponding probability
density of the phases
is the wrapped Gaussian distribution.
Substitution of \eqref{eq:gausans} into Eq.~(\ref{eq:daido})
for $n=1$ yields~\cite{Hannay-Forger-Booth-2018}
\begin{equation}
\dot{Z}_1=(i\omega_0-\gamma)Z_1+h-h^\ast |Z_1|^2Z_1^2-DZ_1\,.
\label{eq:Z1gauss}
\end{equation}

\subsection{Cumulant reduction}
Recently, we suggested~\cite{Tyulkina-etal-2018} a reformulation of the model in terms of the ``circular cumulants'' $\varkappa_n$, instead of the formulation in terms of moments~\eqref{eq:daido}. The cumulants are determined via the power series of the cumulant-gene\-rat\-ing function defined as
\begin{equation}
\Psi(k)=k\frac{\partial}{\partial k}\langle\exp(ke^{i\varphi})\rangle
\equiv\sum_{n=1}^\infty\varkappa_nk^n\,.
\label{eq:cumul}
\end{equation}
For example, the first three circular cumulants are:
$\varkappa_1=Z_1$, $\varkappa_2=Z_2-Z_1^2$, and
$\varkappa_3=(Z_3-3Z_2Z_1+2Z_1^3)/2$.

\begin{figure}[!t]
\centerline{
\includegraphics[width=0.375\textwidth]%
 {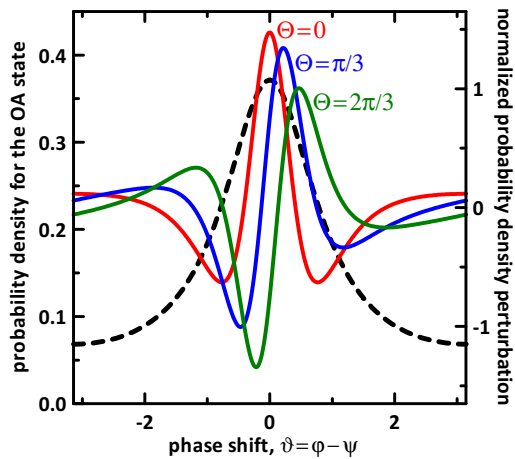}
}
 \caption{The normalized perturbation of the phase
 distribution $w_C/|\varkappa_2|$
 (solid lines, right axis) is compared to the OA distribution $w_{OA}$
 for $|Z_1|=0.4$ (black dashed line, left axis).}
  \label{fig1}
\end{figure}

The merit of the reformulation in terms of the cumulants is two-fold.
\\
(i)~In terms of the cumulants, the OA manifold~\eqref{eq:OAsc}
is a state with one non-vanishing cumulant only: $\varkappa_1=Z_1$ and
$\varkappa_{n>1}=0$. This allows for a representation of the states close
to the OA solution as those with small higher cumulants. The cumulants
$\varkappa_2,\varkappa_3,\ldots$ describe deviations from the OA manifold (from the wrapped
Cauchy distribution), see below Fig.~\ref{fig1} for a visualization
of the perturbation due to $\varkappa_2$.
\\
(ii)~For general states with high synchrony, where
$|Z_1|\approx 1$, the moments $Z_n$ decay slowly with $n$,
while in terms of cumulants one has $|\varkappa_1|\approx 1$ and
$|\varkappa_{n>1}|\ll 1$, which also allows for a nice representation
in terms of cumulants.
In particular, in the case of the wrapped Gaussian distribution~(\ref{eq:gauss}),
the cumulants obey the hierarchy of smallness for arbitrary degree of synchrony;
this hierarchy has a simple analytical form for high and low synchrony:
\begin{equation}
\varkappa_n=
\begin{cases}
\frac{-(-n)^{n-2}}{(n-1)!}e^{in\psi}s^{2(n-1)}\left(1+\mathcal{O}(s^2)\right)
&\textrm{for }s^2\ll 1\,,\\
(-1)^{n-1} Z_1^n\left(1+\mathcal{O}(Z_1^2)\right)&\textrm{for } s^2\gtrsim 1\,.
\end{cases}
\label{eq:hierarchy}
\end{equation}
Hence, the cumulant representation appears to be a proper
framework for perturbations both of the OA solution and of a
highly synchronous state, although the exact equation system for the
cumulants~\cite{Tyulkina-etal-2018} is more complex than~\eqref{eq:daido}:
\begin{equation}
\begin{array}{l}
\dot{\varkappa}_n=n(i\omega_0-\gamma)\varkappa_n+h\delta_{1n}
\\
\qquad
 -h^\ast(n^2{\varkappa}_{n+1}+n\sum\limits_{m=0}^{n-1}\varkappa_{n-m}\varkappa_{m+1})
\\
\qquad\qquad
 -D(n^2{\varkappa}_n+n\sum\limits_{m=0}^{n-2}\varkappa_{n-1-m}\varkappa_{m+1})\,.
\end{array}
\label{eq:kappa}
\end{equation}
Note, that Eqs.~(\ref{eq:daido}) and (\ref{eq:kappa})
describe nonidentical oscillators with Lorentzian distribution
of frequencies; identical ensembles correspond to $\gamma=0$.

In Ref.~\cite{Tyulkina-etal-2018}, the infinite system~(\ref{eq:kappa}) was
analysed in the case of small noise intensity $D$, and
was shown to generate, as a perturbation of the OA
solution, the hierarchy $\varkappa_n\sim D^{n-1}$ for $n\geq 2$.
A first-order correction to the OA ansatz requires
the cumulant $\varkappa_2$ to be taken into account;
from the infinite system~(\ref{eq:kappa}) only two equations remain:
\begin{equation}
\begin{array}{l}
\dot{Z}_1=(i\omega_0-\gamma)Z_1+h-h^\ast(Z_1^2+\varkappa_2)-DZ_1\,,
\\[5pt]
\dot{\varkappa}_2=2(i\omega_0-\gamma)\varkappa_2-4h^\ast(\varkappa_3+Z_1\varkappa_2) -D(4\varkappa_2+2Z_1^2)\,.
\end{array}
\label{eq:Z12c}
\end{equation}
To close these equations, one needs to specify $\varkappa_3$.

The representation of $\varkappa_3$ with maintaining the
first order accuracy can be performed in several ways.
In Ref.~\cite{Tyulkina-etal-2018}, this cumulant was just set to zero:
\begin{equation}
\varkappa_3=0\;.
\label{eq:kappa3=0}
\end{equation}
On the other hand, any substitution $\varkappa_3=\mathrm{const}\,\varkappa_2^2/Z_1$
yields the same first order accuracy for system~(\ref{eq:Z12c}), since it
obeys the hierarchy $\varkappa_n\sim D^{n-1}$.
To find a proper representation of $\varkappa_3$
for a Gaussian distribution with high synchrony,
let us write the first three cumulants for $s\ll 1$: $\varkappa_1=Z_1\approx e^{i\psi}$,
$\varkappa_2\approx -s^2 e^{i2\psi}$,
$\varkappa_3\approx \frac{3}{2}s^4 e^{i3\psi}$. One can see that the closure
\begin{equation}
\varkappa_3=\frac32\frac{\varkappa_2^2}{Z_1}
\label{eq:kappa3-gauss}
\end{equation}
is consistent with this distribution, although it potentially includes
non-Gaussian situations, because in \eqref{eq:kappa3-gauss} $Z_1$
and $\varkappa_2$ are independent of each other. Summarizing,
the closure \eqref{eq:kappa3-gauss} is consistent simultaneously
both with the hierarchy
$\varkappa_n\sim D^{n-1}$ and with the Gaussian distribution with high synchrony,
but generally can describe also states away from these limiting cases.
We stress that the closure~(\ref{eq:kappa3-gauss}) should not be used
in situations, where $Z_1$ is close to zero while $\varkappa_2$ is not small.
For the systems, where $Z_1$ can vanish without $\varkappa_2^2/Z_1$ remaining
finite, a modification to closure~(\ref{eq:kappa3-gauss}) can be suggested:
\begin{equation}
\varkappa_3=\frac32\varkappa_2^2Z_1^\ast\,;
\label{eq:kappa3-gauss2}
\end{equation}
this modification is equivalent to Eq.~(\ref{eq:kappa3-gauss}) at $s\to0$,
but less accurately corresponds to the wrapped Gaussian distribution for $|Z_1|<1$.
It is also not less accurate than the first-order correction to the OA solution.

It is instructive to visualize the perturbation of the OA probability
density corresponding to one nonvanishing second circular
cumulant $\varkappa_2$. With two nonvanishing cumulants,
the moment-generating function is
\[
F(k)=\sum_{m=0}^{\infty}Z_m(t)\frac{k^m}{m!}=
\exp\Big[kZ_1+\varkappa_2\frac{k^2}{2}\Big]\;.
\]
Assuming smallness of $\varkappa_2$, we approximate
it as $F(k)\approx (1+\varkappa_2\frac{k^2}{2})\exp[kZ_1]$,
and obtain for the moments $Z_m=Z_1^m+\frac{m(m-1)}{2}\varkappa_2 Z_1^{m-2}$.
Summation of the Fourier series with these Fourier
coefficients yields $w(\varphi)=w_{OA}(\varphi)+w_C(\varphi)$, where
\[
w_{OA}(\varphi)=\frac{1-|Z_1|^2}{2\pi|e^{i\varphi}-Z_1|^2}
\]
is the wrapped Cauchy distribution corresponding to the OA ansatz, and
\[
w_C(\varphi)=\text{Re}\!\left[\frac{\pi^{-1}\varkappa_2 e^{i\varphi}}{\left(e^{i\varphi}-Z_1\right)^3}\right].
\]
is the correction corresponding to a nonvanishing second cumulant.
We illustrate the perturbation of the probability density in Fig.~\ref{fig1}.
We depict the OA-density relative to the argument of the order parameter,
by using $\vartheta=\varphi-\text{arg}(Z_1)$. One can see that
the perturbation is localized close to the maximum of
the unperturbed density $w_{OA}$; its exact position depends
on the difference of the arguments of the two cumulants
involved $\Theta=\text{arg}(\varkappa_2)-2\text{arg}(Z_1)$.

\section{Accuracy of different approximations}
Above we have outlined five possible finite-di\-men\-sional descriptions
of the noisy interacting
population: Eqs.~(\ref{eq:Z1OA}), (\ref{eq:Z1gauss}),
(\ref{eq:Z12c},\ref{eq:kappa3=0}), (\ref{eq:Z12c},\ref{eq:kappa3-gauss}), and
(\ref{eq:Z12c},\ref{eq:kappa3-gauss2}) (cf. Table~\ref{tab1} below).
The OA equation~\eqref{eq:Z1OA} is exact for noiseless populations $D=0$.
The Gaussian ansatz~\eqref{eq:Z1gauss} is exact for noisy oscillators without
coupling $h=0$. The two-cumulant (2C) approximations~(\ref{eq:Z12c})
with closures~(\ref{eq:kappa3=0}), (\ref{eq:kappa3-gauss}), or (\ref{eq:kappa3-gauss2})
reduce to the OA ansatz for $D=0$ if one sets $\varkappa_2=0$,
and in fact are the first-order corrections in the noise
intensity $D$; we will use them, however, for large values of $D$ as well.

For the ensemble of identical oscillators ($\gamma=0$)
in a steady state, where $h=const$ (in a rotating reference frame,
if necessary), the stationary distribution of phases according
to \eqref{eq:fpf} is the von Mises distribution
\[
w=\frac{\exp\left[\frac{2|h|}{D}
\cos\big(\varphi-\text{arg}(h)\big)\right]}{2\pi I_0(2|h|/D)}\,,
\]
where $I_0(\cdot)$ is the modified Bessel function of order $0$.
In the case $|h|\gg D$, it is close to the wrapped Gaussian distribution;
thus, one expects that the Gaussian approximation will provide an
accurate steady state in this limit. Simultaneously, this is the
case of high synchrony, where substitutions~(\ref{eq:kappa3-gauss}) and
(\ref{eq:kappa3-gauss2}) are relevant.

Below we compare the accuracy of the steady states
according to the approximations outlined,
for the Kuramoto model~\eqref{eq:KM} and the
active rotator model~\eqref{eq:ARM}.

\subsection{Kuramoto model}
The Kuramoto model for noisy oscillators~\eqref{eq:KM} contains
three parameters: $\gamma$, $D$, and $K$. However, by virtue of
a time normalization, one can get rid of one parameter. The
critical coupling for the onset of synchronization is
$K_\mathrm{cr}=2(\gamma+D)$. Thus, it is convenient to
choose $\gamma=1-D$, so that the critical
coupling is $K_\mathrm{cr}=2$.

\begin{figure}[!t]
\center{
\includegraphics[width=0.98\columnwidth]%
 {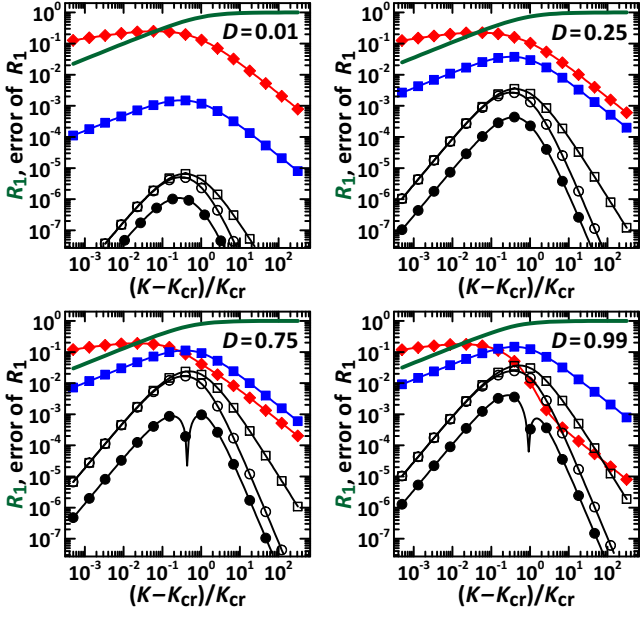}
}
\caption{The accuracy of solutions for the noisy Kuramoto
ensemble {\it vs} coupling strength $K$ is plotted with
blue solid squires for the Ott-Antonsen ansatz~(\ref{eq:KM-OA}),
red solid diamonds for the Gaussian approximation~(\ref{eq:KM-gauss}),
black open squares for the 2C truncation with closure $\varkappa_3=0$~(\ref{eq:KM-2c0}),
black open circles for closure $\varkappa_3=1.5\varkappa_2^2Z_1^\ast$,
black solid circles for closure $\varkappa_3=1.5\varkappa_2^2/Z_1$ (\ref{eq:KM-2c})
(the cusp at a large noise strength is due to the change of sign of the error).
Bold solid lines: the exact solution for the order parameter $R_1$.
Parameters are rescaled so that $\gamma+D=1$: noise intensity $D$
is specified in plots, $K_\mathrm{cr}=2(\gamma+D)=2$.
The case of vanishing intrinsic noise corresponds to $D=0$,
the case of identical natural frequencies (or extremely
strong noise) corresponds to $D=1$.
}
  \label{fig2}
\end{figure}

First, we calculated the ``exact'' steady state of system~(\ref{eq:kappa})
by solving it with 200 cumulants taken into account.
Then we found the steady solutions of approximations~(\ref{eq:Z1OA}), (\ref{eq:Z1gauss}), (\ref{eq:Z12c},\ref{eq:kappa3=0}),
(\ref{eq:Z12c},\ref{eq:kappa3-gauss}):
\begin{align}
&
R_1^2=1-\frac{K_\mathrm{cr}}{K}\,,
\label{eq:KM-OA}
\\
&
R_1^2=\sqrt{1-\frac{K_\mathrm{cr}}{K}}\,,
\label{eq:KM-gauss}
\\
&
R_1^2=\frac{1}{2}-\frac{3K_\mathrm{cr}}{4K}
+\frac{\sqrt{(2K-K_\mathrm{cr})^2+16D(K-K_\mathrm{cr})}}{4K}\,,
\label{eq:KM-2c0}
\\
&
R_1^2=2-\frac{3K_\mathrm{cr}}{2K}
-\frac{\sqrt{4(K-K_\mathrm{cr})(K-2D)+K_\mathrm{cr}^2}}{2K}\,,
\label{eq:KM-2c}
\end{align}
respectively; the approximation~(\ref{eq:Z12c},\ref{eq:kappa3-gauss2})
yields a cubic equation for $R_1^2$ with a cumbersome analytical solution.
The deviations from the exact state are shown in Fig.~\ref{fig2}.
One can see, that the Gaussian approximation yields better accuracy
than the OA ansatz only for strong noise ($D=0.99$,
we remind that according to the
normalization adopted $0\leq D\leq 1$)
and strong coupling.
The 2C approximation with the closure $\varkappa_3=0$ works as a plain first-order
correction to the OA solution.
The 2C approximation with the closure $\varkappa_3=(3/2)\varkappa_2^2/Z_1$ (\ref{eq:kappa3-gauss})
is the best one in all situations.
The 2C approximation with the closure $\varkappa_3=(3/2)\varkappa_2^2Z_1^\ast$ (\ref{eq:kappa3-gauss2})
is approaching the one with (\ref{eq:kappa3-gauss}) for high synchrony, but yields the same
accuracy as the closure $\varkappa_3=0$ for small $R_1$; for a strong noise and moderate
synchrony, it is only slightly less accurate than the Gaussian approximation.
Close to the synchronization threshold $K_\mathrm{cr}$, the inaccuracy
of the Gaussian approximation reaches $0.2$ and exceeds the
value of order parameter $R_1$, while the inaccuracy of $R_1$
with the OA ansatz is always reasonably small.

As the 2C approximations are based on the correction to the
OA one, the former are always superior to the latter. One can also see,
that the Gaussian approximation is accurate where the synchrony is
high, which is also suggested by the von Mises distribution with $|h|\gg D$.
Noteworthy, for high synchrony, the 2C approximations with
closures~(\ref{eq:kappa3-gauss}) and (\ref{eq:kappa3-gauss2})
contain the Gaussian distribution
as an admissible particular case. Moreover, these 2C truncations
employ the Gaussian scaling only in the expression for the third
cumulant $\varkappa_3$,
while the second cumulant $\varkappa_2$ is allowed to deviate
from the value dictated by the first cumulant $Z_1$
for the Gaussian distribution. Hence, these truncations also encompass
a first-order correction for the case of Gaussian approximation under
high synchrony. The closure~(\ref{eq:kappa3-gauss}) decently approximates
the wrapped Gaussian distribution also for non-high synchrony. Being not
less accurate than the first-order corrections
to both the OA and Gaussian reductions, the 2C reduction
with closure~(\ref{eq:kappa3-gauss}) becomes
superior to them for the Kuramoto model with intrinsic noise.

\subsection{Active rotators model}

\begin{figure}[!t]
\center{
\includegraphics[width=0.98\columnwidth]%
 {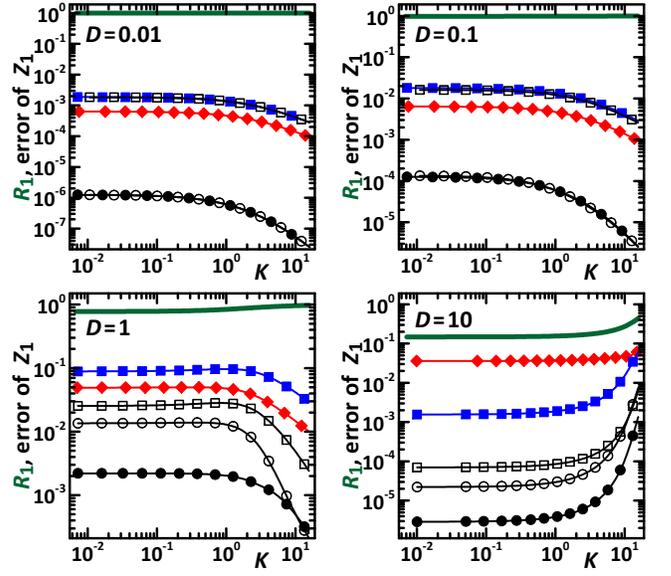}
}
\caption{The difference between the exact
steady-state solution for the population of noisy
active rotators with Ku\-ra\-mo\-to-type coupling~(\ref{eq:ARM})
and different approximations {\it vs}
the coupling strength $K$.
Blue solid squares: the OA ansatz~(\ref{eq:Z1OA});
red solid diamonds: the Gaussian ansatz~(\ref{eq:Z1gauss});
black open squares: the two-cumulant reduction~(\ref{eq:Z12c}) with closure~(\ref{eq:kappa3=0});
black open circles: reduction~(\ref{eq:Z12c}) with~(\ref{eq:kappa3-gauss2});
black solid circles: reduction~(\ref{eq:Z12c}) with~(\ref{eq:kappa3-gauss}).
Bold solid lines: the order parameter $R_1$ for the accurate solution.
Parameters: $\omega_0=1$, $a=3$, noise intensity $D$ is specified in the panels.
}
  \label{fig3}
\end{figure}

A population of active rotators with the Kuramoto-type coupling~\eqref{eq:ARM}
can exhibit diverse regimes of collective dynamics, depending on
parameter values~\cite{Zaks-etal-2003}.
Following Ref.~\cite{Zaks-etal-2003}, we consider identical
elements ($\gamma=0$) and focus on the case which is
impossible for the Kuramoto ensemble: an excitable state of
individual elements, $a>\omega_0$. Noteworthy, in this case
the synchrony imperfectness is owned solely by intrinsic
Gaussian noise. For all the cases presented in Figs.~\ref{fig3}
and \ref{fig4}, the accurate solution is calculated
from system~(\ref{eq:kappa}) with $200$ cumulants.
In Fig.~\ref{fig3}, we evaluate accuracy of
approximations~(\ref{eq:Z1OA}), (\ref{eq:Z1gauss}), (\ref{eq:Z12c},\ref{eq:kappa3=0}), (\ref{eq:Z12c},\ref{eq:kappa3-gauss}), and (\ref{eq:Z12c},\ref{eq:kappa3-gauss2}).
For high synchrony (which is observed for a weak noise), the
Gaussian approximation is more accurate than the OA one.
Where the OA approximation fails, the plain 2C approximation with
closure $\varkappa_3=0$ is not more accurate than the OA solution:
in Fig.~\ref{fig3} for $D=0.01$ and $0.1$, the inaccuracy of the
OA solution is of the same order of magnitude as the deviation of $R_1$ from $1$ for the exact solution.
The 2C approximations with Gaussian closures always provide
much better accuracy than both the OA and the Gaussian ones.

\subsection{Testing scaling laws}

\begin{figure}[!t]
\center{
\includegraphics[width=0.999\columnwidth]%
 {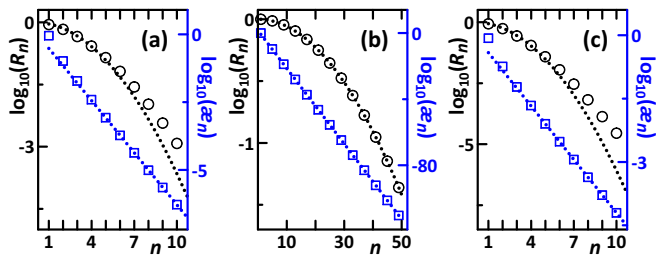}
}
\caption{The scaling law for the Kuramoto-Diado order parameters
$R_n$ (circles) and the hierarchy of smallness for the circular
cumulants $\varkappa_n$ (squares) are plotted for the noisy
Kuramoto system with $D=0.75$, $\gamma=0.25$, $K=3K_\mathrm{cr}$~(a)
and the population of active rotators with $\omega=1$, $a=3$, $K=1$,
$\gamma=0$, $D=0.01$~(b) and $D=1$~(c). The solutions of
system~(\ref{eq:kappa}) are calculated with $200$ cumulants.
Dotted lines show the trends $R_n\sim R_1^{n^2}$ and
$\varkappa_n\propto{s^{2(n-1)}}$.
}
  \label{fig4}
\end{figure}

In Fig.~\ref{fig4} we test how well the scaling laws~\eqref{eq:OAsc}
and~\eqref{eq:gausans}, which lie at the basis of the OA and the
Gaussian approximations, are valid. To check the OA ansatz~\eqref{eq:OAsc},
we plot the values of the cumulants: the cumulants $\varkappa_n$ with $n\geq 2$
should vanish if the OA ansatz is exact. To check the Gaussian approximation,
we compare the $n$-dependence of $R_n=|Z_n|$ with a parabola.

Panel Fig.~\ref{fig4}(a) shows the scaling for the Kuramoto model.
One can see that although high-order cumulants do not vanish,
there is a gap between the first and the second cumulants.
This means that the OA ansatz is relatively good, but can be definitely
improved by taking into account the second cumulant. The Gaussian approximation
is valid for small $n\lesssim 7$ only.

Panels Fig.~\ref{fig4}(b,c) show the cumulants and the moments
for the active rotator model. In panel~(b) we illustrate the situation where
the Gaussian approximation is superior to the OA one. One can see that
the system practically perfectly obeys the $n^2$-scaling law for $R_n$. On the other
hand, the gap between the first and the second cumulants is not large,
which means that the OA ansatz is poor (see Fig.~\ref{fig3} for $D=0.01$, $K=1$);
the inaccuracy of the OA solution is
compatible to the deviation of $R_1$ from $1$. The case
in panel~(b) is the case of high synchrony. The plots
in panel (c) are similar to those in panel~(a); here only
a few first values $R_n$ follow the $n^2$-scaling law~\eqref{eq:gausans}.
On the other hand, the gap between the first and the second cumulants is present,
and the OA ansatz becomes acceptably accurate (see Fig.~\ref{fig3} for $D=1$, $K=1$);
here the inaccuracy of the OA solution is one order of magnitude smaller
than the deviation of $R_1$ from $1$.

Remarkably, in all the cases one observes that higher cumulants
decay exponentially $\varkappa_n\propto \exp[-\mathrm{const}\,n]$.
This law has been derived in Ref.~\cite{Tyulkina-etal-2018} for
small $D$ only; here we see that it is valid for moderate and
strong noise as well.
For a strong noise, there is a small parameter $(1/D)$
which can serve for a hierarchy in the system,
$\varkappa_{n+1}\sim(1/D)\varkappa_n$. For a
moderate noise strength, there is no small parameter,
but nevertheless, a hierarchy is present.

\begin{table}[t]
\caption{Low-dimensional model reductions}
\begin{center}
\begin{tabular}{p{2.1cm}p{1.15cm}p{4.9cm}}
\hline\hline
\centerline{Reduction} & \centerline{Eqs.} & \centerline{Comments}
 \\
\hline
\multicolumn{2}{p{3.25cm}}{$\begin{tabular}{p{2.1cm}p{1.15cm}}
OA ansatz
 & \centerline{(\ref{eq:Z1OA})}
\\[5pt]
Gaussian approximation
 & \centerline{(\ref{eq:Z1gauss})}
\end{tabular}$} &
{$\begin{tabular}{p{4.9cm}}
The Gaussian approximation is superior to the OA one for high synchrony,
if the distortion of the perfect synchrony is not dominantly due to a
non-Gaussian disorder (e.g.\ Lorentzian distribution of natural frequencies).
\\[5pt]
\end{tabular}$}
  \\[5pt]
Two-cumulant reduction with $\varkappa_3=0$
 & \centerline{(\ref{eq:Z12c},\ref{eq:kappa3=0})} &
This plain first-order correction to the OA solution is frequently superior
to the Gaussian approximation, but may have the same (low) accuracy as the OA
ansatz, where the latter completely fails.
  \\[5pt]
Two-cumulant reduction with $\varkappa_3=\frac32\varkappa_2^2/Z_1$
 & \centerline{(\ref{eq:Z12c},\ref{eq:kappa3-gauss})} &
It works as a first-order correction to the best of OA and Gaussian approximations.
Not to be used for problems where $Z_1$ can approach $0$ without $\varkappa_2^2/Z_1$
remaining finite.
  \\[5pt]
Two-cumulant reduction with $\varkappa_3=\frac32\varkappa_2^2Z_1^\ast$
 & \centerline{(\ref{eq:Z12c},\ref{eq:kappa3-gauss2})} &
For high synchrony, its accuracy approaches the accuracy of
closure $\varkappa_3=\frac32\varkappa_2^2/Z_1$. For low synchrony,
it is as accurate as the plain two-cumulant truncation ($\varkappa_3=0$).
  \\[3pt]
\hline\hline
\end{tabular}
\end{center}
\label{tab1}
\end{table}

\section{Conclusion}
We have compared five low-dimensional approximations
describing the dynamics of large populations
of noisy phase oscillators (or active rotators)
with global sine-coupling: Eqs.~(\ref{eq:Z1OA}), (\ref{eq:Z1gauss}),
(\ref{eq:Z12c},\ref{eq:kappa3=0}), (\ref{eq:Z12c},\ref{eq:kappa3-gauss}),
and (\ref{eq:Z12c},\ref{eq:kappa3-gauss2}); the latter two cases are novel two-cumulant
truncations within the framework
of circular cumulant formalism.
As prototypic examples, we have chosen the standard Kuramoto model and
the active rotator model in the excitable state regime.
Tabel~\ref{tab1} summarizes applicability of different low-dimensional reductions.
The truncation with the closure according to
$\varkappa_3=(3/2)\varkappa_2^2/\varkappa_1$, which most accurately corresponds
to the Gaussian reduction under high synchrony, deserves special attention. By construction,
this truncation is simultaneously a first-order correction to
the Ott-Antonsen ansatz, and comprises the wrapped Gaussian
distribution of phases, where the latter can be formed. In all the cases
considered, this two-cumulant approximation is
significantly superior to all other approximations.
Remarkably, even for the cases, where $R_n$ nearly
perfectly follows the $n^2$-scaling law, this two-cumulant
approximation enhances the accuracy of the Gaussian
one, by a few orders of magnitude.

Generally, a high synchrony is not a sufficient
condition for applicability of the Gaussian ansatz.
In this paper, our analysis has been restricted to
the situations of synchronization by coupling. However,
for synchronization by a
common noise~\cite{Pikovsky-1984b,Teramae-Tanaka-2004,Goldobin-Pikovsky-2004},
in nonideal situations (i.e., with
intrinsic noise and/or nonidentity of elements),
it is known that the phase distribution possesses
heavy power-law tails even in the limit of high
synchrony~\cite{Goldobin-Pikovsky-2005b,Pimenova-etal-2016}.
For such systems, the Gaussian ansatz is never natural.

\acknowledgments{
The authors thank M.\ Zaks for fruitful discussions
and Z.\ Levnajic for bringing paper~\cite{Hannay-Forger-Booth-2018}
to our attention.
Work of A.P.\ on Secs.~II, III, V.C was supported by
Russian Science Foundation (Grant Nr.\ 17-12-01534).
Work of D.S.G.\ and L.S.K.\ on Secs.~IV, V.A-B was
supported by Russian Science Foundation (Grant Nr.\ 14-21-00090).}

\end{document}